\documentclass{webofc}
\usepackage[varg]{txfonts}
\usepackage{hyperref}
\usepackage{url}
\hypersetup{colorlinks=true,citecolor=blue,urlcolor=blue,linkcolor=blue}
\begin{document}
\title{Cryogenic Systems for the TUCAN EDM Experiment}

\author{
\firstname{Jeffery W.} \lastname{Martin}\inst{1}\fnsep\thanks{\email{j.martin@uwinnipeg.ca}}\fnsep\thanks{on behalf of the TUCAN Collaboration}
\firstname{B.} \lastname{Algohi}\inst{2}\and
\firstname{D.} \lastname{Anthony}\inst{3}\and 
\firstname{L.} \lastname{Barr\'on-Palos}\inst{4}\and
\firstname{M.} \lastname{Bradley}\inst{5}\and
\firstname{A.} \lastname{Brossard}\inst{3}\and
\firstname{T.} \lastname{Bui}\inst{2}\and
\firstname{J.} \lastname{Chak}\inst{3}\and
\firstname{C.} \lastname{Davis}\inst{3}\and
\firstname{R.} \lastname{de Vries}\inst{1}\and 
\firstname{K.} \lastname{Drury}\inst{3}\and 
\firstname{D.} \lastname{Fujimoto}\inst{3}\and
\firstname{R.} \lastname{Fujitani}\inst{6}\and
\firstname{M.} \lastname{Gericke}\inst{2}\and
\firstname{P.} \lastname{Giampa}\inst{3}\and
\firstname{R.} \lastname{Golub}\inst{7}\and
\firstname{T.} \lastname{Hepworth}\inst{1}\and 
\firstname{T.} \lastname{Higuchi}\inst{8,9}\and
\firstname{G.} \lastname{Ichikawa}\inst{10}\and
\firstname{S.} \lastname{Imajo}\inst{9}\and
\firstname{A.} \lastname{Jaison}\inst{2}\and
\firstname{B.} \lastname{Jamieson}\inst{1}\and
\firstname{M.} \lastname{Katotoka}\inst{1}\and
\firstname{S.} \lastname{Kawasaki}\inst{10}\and
\firstname{M.} \lastname{Kitaguchi}\inst{11}\and
\firstname{W.} \lastname{Klassen}\inst{12}\and
\firstname{E.} \lastname{Korkmaz}\inst{13}\and
\firstname{E.} \lastname{Korobkina}\inst{7}\and
\firstname{M.} \lastname{Lavvaf}\inst{2}\and
\firstname{T.} \lastname{Lindner}\inst{3,1}\and
\firstname{N.} \lastname{Lo}\inst{3}\and 
\firstname{S.} \lastname{Longo}\inst{2}\and
\firstname{K.} \lastname{Madison}\inst{12}\and
\firstname{Y.} \lastname{Makida}\inst{10}\and
\firstname{J.} \lastname{Malcolm}\inst{3}\and 
\firstname{J.} \lastname{Mammei}\inst{2}\and
\firstname{R.} \lastname{Mammei}\inst{1}\and
\firstname{C.} \lastname{Marshall}\inst{3}\and
\firstname{M.} \lastname{McCrea}\inst{1}\and
\firstname{E.} \lastname{Miller}\inst{12}\and
\firstname{M.} \lastname{Miller}\inst{14}\and
\firstname{K.} \lastname{Mishima}\inst{11}\and
\firstname{T.} \lastname{Mohammadi}\inst{2}\and
\firstname{T.} \lastname{Momose}\inst{12}\and
\firstname{T.} \lastname{Okamura}\inst{10}\and
\firstname{H.J.} \lastname{Ong}\inst{9}\and
\firstname{R.} \lastname{Patni}\inst{3}\and 
\firstname{R.} \lastname{Picker}\inst{3,15}\and
\firstname{W.D.} \lastname{Ramsay}\inst{3}\and
\firstname{W.} \lastname{Rathnakela}\inst{2}\and
\firstname{J.} \lastname{Sato}\inst{11}\and
\firstname{W.} \lastname{Schreyer}\inst{3,16}\and
\firstname{T.} \lastname{Shima}\inst{9}\and
\firstname{H.} \lastname{Shimizu}\inst{11}\and
\firstname{S.} \lastname{Sidhu}\inst{3}\and
\firstname{S.} \lastname{Stargardter}\inst{2}\and
\firstname{P.} \lastname{Switzer}\inst{1}\and 
\firstname{I.} \lastname{Tanihata}\inst{9}\and
\firstname{S.} \lastname{Vanbergen}\inst{12}\and
\firstname{W.T.H.} \lastname{van~Oers}\inst{2,3}\and
\firstname{Y.} \lastname{Watanabe}\inst{10}\and
\firstname{A.} \lastname{Zahra}\inst{2}\and
\firstname{M.} \lastname{Zhao}\inst{3} 
}

\institute{
  The University of Winnipeg, Winnipeg, MB, Canada\and
  University of Manitoba, Winnipeg, MB, Canada\and
  TRIUMF, Vancouver, BC, Canada\and
  Universidad Nacional Aut\'onoma de M\'exico, Mexico City, Mexico\and
  University of Saskatchewan, Saskatoon, SK, Canada\and
  Department of Nuclear Engineering, Kyoto University, Kyoto, Japan\and
  North Carolina State University, Raleigh, NC, USA\and
  Institute for Integrated Radiation and Nuclear Science (KURNS), Kyoto University, Osaka, Japan\and
  Research Center for Nuclear Physics (RCNP), Osaka University, Osaka, Japan\and
  High Energy Accelerator Research Organization (KEK), Tsukuba, Ibaraki, Japan\and
  Nagoya University, Aichi, Japan\and
  The University of British Columbia, Vancouver, BC, Canada\and
  The University of Northern BC, Prince George, BC, Canada\and
  McGill University, Montreal, QC, Canada\and
  Simon Fraser University, Burnaby, BC, Canada\and
  Oak Ridge National Laboratory, Knoxville, TN, USA
}

\abstract{

  The TUCAN (TRIUMF UltraCold Advanced Neutron) Collaboration is
  completing a new ultracold neutron (UCN) source.  The UCN source
  will deliver UCNs to a neutron electric dipole moment (EDM)
  experiment.  The EDM experiment is projected to be capable of an
  uncertainty of $1\times 10^{-27}~e$cm, competitive with other
  planned projects, and a factor of ten more precise than the present
  world's best.  The TUCAN source is based on a UCN production volume
  of superfluid helium (He-II), held at 1~K, and coupled to a
  proton-driven spallation target.  The production rate in the source
  is expected to be in excess of $10^7$~UCN/s; since UCN losses can be
  small in superfluid helium, this should allow us to build up a large
  number of UCNs.  The spallation-driven superfluid helium technology
  is the principal aspect making the TUCAN project unique.  The
  superfluid production volume was recently cooled, for the first
  time, and successfully filled with superfluid helium.  The design
  principles of the UCN source are described, along with some of the
  challenging cryogenic milestones that were recently passed.
  
}

\maketitle

\section{Scientific motivation:  the neutron electric dipole moment}
\label{intro}

The neutron electric dipole moment (nEDM) is an experimental
observable of high importance in fundamental physics because it
violates time-reversal symmetry and therefore CP (charge-parity)
symmetry~\cite{bib:pospelov,bib:engel,bib:chuppall}, the symmetry
relating the interactions of particles to those of their antiparticle
counterparts. To date, all experiments have found the nEDM to be
compatible with zero.  Improving the experimental precision places
tighter constraints on new sources of CP violation beyond the Standard
Model.  Conversely, if a small but non-zero nEDM were discovered, it
would herald a discovery of new physics.  Even if ascribed to the
CP-violating $\bar{\theta}$ parameter of the strong sector, the
mystery of a small but non-zero $\bar{\theta}$ would create a new
problem for the Standard Model.

A recent measurement performed using ultracold neutrons (UCNs) at the
Paul Scherrer Institute (PSI) determined an upper bound on the nEDM,
$|d_n|<1.8\times 10^{-26}~e$cm (90\% C.L.)~\cite{bib:psi2020}.  In
addition to setting a new world record in precision, this work is
noteworthy in that it is the first nEDM measurement conducted using a
superthermal UCN source, a strategy pursued by our project and a host
of new UCN sources that are expected to revolutionize the field.

Recent theoretical work addressing the physics impact of an even more
precise measurement of the nEDM has focused on three general (and
overlapping) themes: (1) new sources of CP violation beyond the
Standard Model~\cite{bib:cirigliano,bib:crivellin}, (2) baryogenesis
scenarios, especially new physics contributions to electroweak
baryogenesis inspired scenarios~\cite{bib:bell,bib:hou} and (3) the
strong CP problem, related to searches for
axions~\cite{bib:carena,bib:mimura,bib:peinado,bib:psiaxion}.  Because
of these connections, better measurements of the nEDM are of vital
importance in particle physics and early universe cosmology.

Next generation UCN EDM experiments are in preparation at a variety of
sites, worldwide, and are aiming to improve the result by an order of
magnitude or more.  Experiments are planned at Institut Laue-Langevin
(ILL, Grenoble, France)~\cite{bib:panedm}, PSI~\cite{bib:n2edm}, and
Los Alamos National Laboratory (LANL, Los Alamos, NM,
USA)~\cite{bib:alarcon,bib:lanledm}, and include our effort at
TRIUMF~\cite{bib:npn}.  Our goal of $\sigma(d_n)<10^{-27}~e$cm is
competitive with these efforts.

The TRIUMF UltraCold Neutron (TUCAN) Collaboration represents a
collaboration of physicists from Canada, Japan, M\'exico, and the
United States.  Our experiment is conducted at TRIUMF, Canada's
Particle Accelerator Centre, located in Vancouver, British Columbia.
It uses a high-energy, high-intensity proton beam from the TRIUMF
cyclotron impinging on a water-cooled tungsten target to initiate
spallation reactions which liberate neutrons.

The basis of the TUCAN approach involves a spallation-driven,
superfluid $^4$He UCN source connected to a room-temperature EDM
experiment.  The key component making our project unique is our UCN
source which, after the completion of our upgrade, is expected to
become the world's best.  We envision achieving UCN counting rates
over 100 times larger than the last experiment at PSI, similar to or
surpassing the plans of other experiments, and enabling the next
breakthrough for this field.

\section{Experiment overview}
\label{expt}

\begin{figure}[h]
\begin{center}
\includegraphics[width=\textwidth]{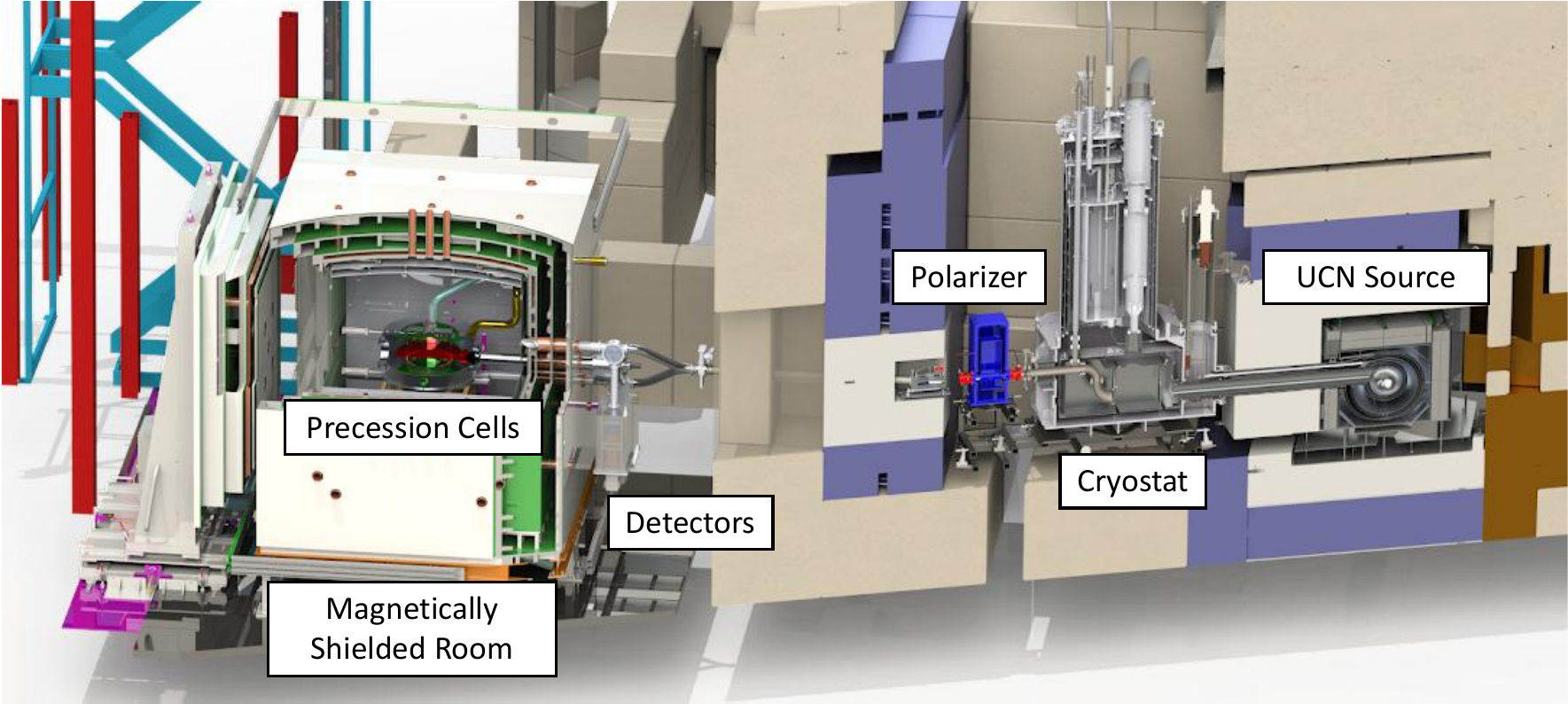}
\caption{UCN source and EDM spectrometer for the TUCAN project.  UCN
  exiting the source are polarized by a superconducting magnet and
  pass through UCN guides to reach the EDM experiment located within
  the MSR.  UCN spins precess in dual EDM measurement cells with a
  holding field provided by a $B_0$ coil and electric field provided
  by a central HV electrode. UCN spin analyzers sense the neutron
  spins at the end of each cycle.}
\label{fig:MesonCAD}
\end{center}
\end{figure}

Figure~\ref{fig:MesonCAD} shows a schematic diagram of the planned
apparatus.  At the present time, all components of the UCN source have
been completed, except for the liquid deuterium cryostat.  The UCN
source itself will be described in further detail in
Section~\ref{ucnsource}.  All components of the nEDM experiment have
been prototyped and final versions are being built.  The magnetically
shielded room (MSR) needed for the nEDM experiment has been completed
and has been shown to meet the specifications needed for a
$10^{-27}~e$cm measurement.

\begin{figure}[h]
\centering
\includegraphics[width=\textwidth]{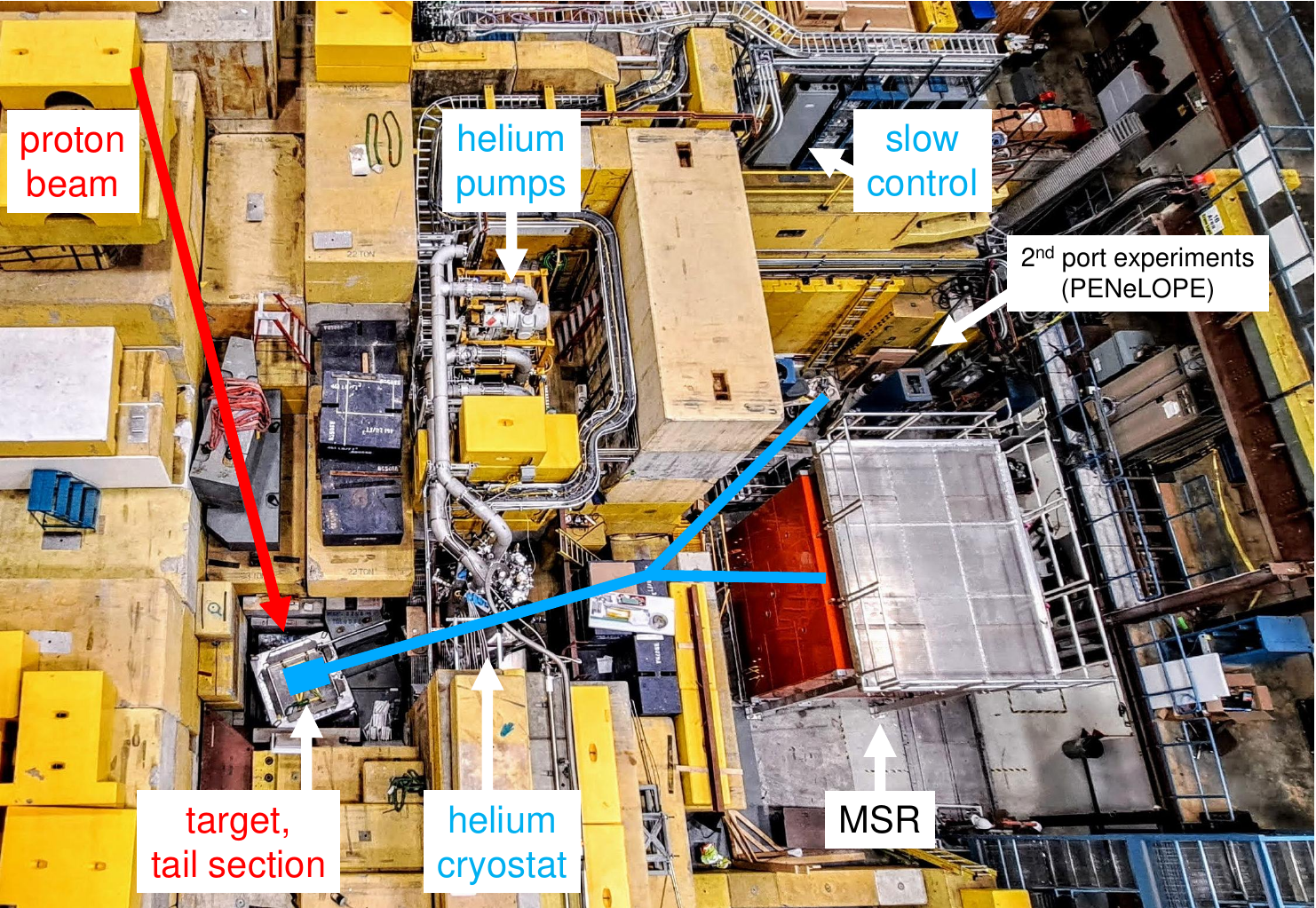}
\caption{Overhead view of the UCN source facility (April 2024).  Lines
  display the underlying proton beam path (red) and sketch the
  existing and planned UCN guide paths (blue).\label{fig:overhead}}
\end{figure}

An overhead view of the facility is shown in Fig.~\ref{fig:overhead}.
The photograph was taken in April 2024, just before the UCN source was
covered in shielding blocks.  The facility is now complete to the
stage that the UCN source has been operated in a month-long cryogenic
test run, and the magnetically shielded room (MSR) is routinely in use
for magnetometer and coil testing, in preparation for UCN runs planned
for late 2025.

\section{Ultracold neutron source principles and design}
\label{ucnsource}

Our UCN source is based on previous work reported in
Refs.~\cite{bib:masuda1,bib:masuda2}.  The prototype vertical UCN
source developed in Japan was moved to Canada and installed in the
Meson Hall at TRIUMF in 2017.  We developed and constructed a new
spallation target and proton beamline at TRIUMF for operation up to
40~$\mu$A (in preparation for an upgraded UCN
source)~\cite{bib:ucnbeamline}.  The beamline features a fast kicker
system which allows us to run simultaneously with other Meson Hall
users~\cite{bib:kicker}.  We did experiments with the vertical source
on UCN production~\cite{bib:physrevc}, transport and
storage~\cite{bib:moderators}, and polarization and
detection~\cite{bib:shrthesis}.  In 2020-21, the vertical source was
decommissioned in preparation for the upgrade.

\begin{figure}[h]
\centering
\includegraphics[width=\textwidth]{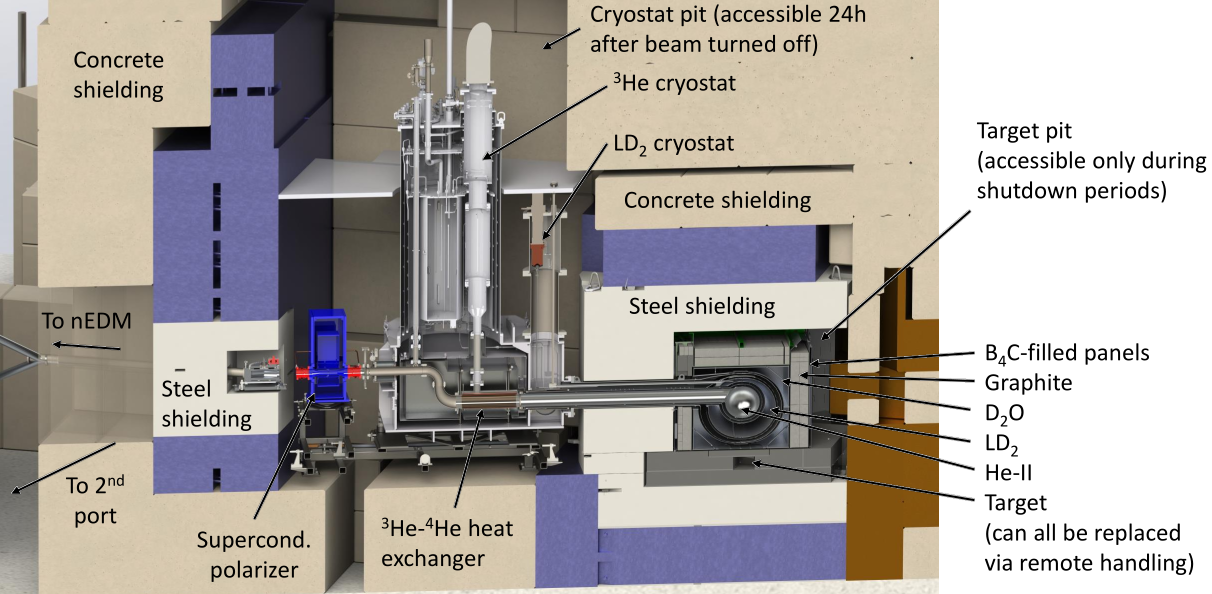}
\caption{The recently completed UCN source.  Neutrons are liberated by
  proton-induced spallation at 480~MeV and 40~$\mu$A in a target
  located beneath the He-II, LD$_2$, and D$_2$O volumes.  Neutrons are
  reflected and moderated in surrounding materials then enter
  superfluid $^4$He (He-II) where they are downscattered to become
  UCNs.  UCNs created in the He-II are transported out through the
  heat exchanger passing through the superconducting polarizer magnet
  to the nEDM experiment.\label{fig:UCNsourceCAD}}
\end{figure}

The new UCN source upgrade features a 27~L UCN production volume (the
He-II volume in Fig.~\ref{fig:UCNsourceCAD}) which is significantly
larger than the 8~L volume used in the vertical source.  The vertical
source could reliably handle 300~mW of heat load to the He-II whereas
the horizontal source is optimized to handle
10~W~\cite{bib:shinsuke,bib:okamura}.  A new, larger capacity helium
pumping system enables the additional cooling power.  Additionally, a
new large-area $^3$He-$^4$He heat exchanger
(Fig.~\ref{fig:UCNsourceCAD}) was built to be compatible with both UCN
transport and heat transfer requirements, resolving a severe
limitation of the vertical source~\cite{bib:shrthesis}.

We completed detailed estimates for UCN production and
extraction~\cite{bib:moderators,bib:ss,bib:ssthesis} based on a Monte
Carlo N-Particle (MCNP) model of the source, a model of UCN production
based on Ref.~\cite{bib:korobk}, and UCN transport simulations based
on Ref.~\cite{bib:pentrack} including losses within the He-II and in
transport to the EDM experiment.  The optimization indicates that when
driven by a 40~$\mu$A proton beam, the source will produce $1.4\times
10^7$~UCN/s, with beam heating of 8.1~W to the He-II at 1.1~K. This is
more than two orders of magnitude larger than the UCN production rate
of the vertical source.  A total of $1.38\times 10^7$~UCN would be
loaded into the EDM measurement cells prior to initiating the Ramsey
cycle.  Using reasonable values for lifetimes and spin-coherence times
of the UCN, this corresponds to a statistical determination of the
nEDM of $\sigma(d_n)=3\times 10^{-25}~e$cm per cycle.  Using
conservative assumptions for the running time available per day, a
statistical determination of $\sigma(d_n)=10^{-27}~e$cm would be
achieved within 280~days of running~\cite{bib:ss}.

\section{Recent cryogenic performance results}
\label{cryo}

\begin{figure}[h]
\begin{center}
\includegraphics[width=\textwidth]{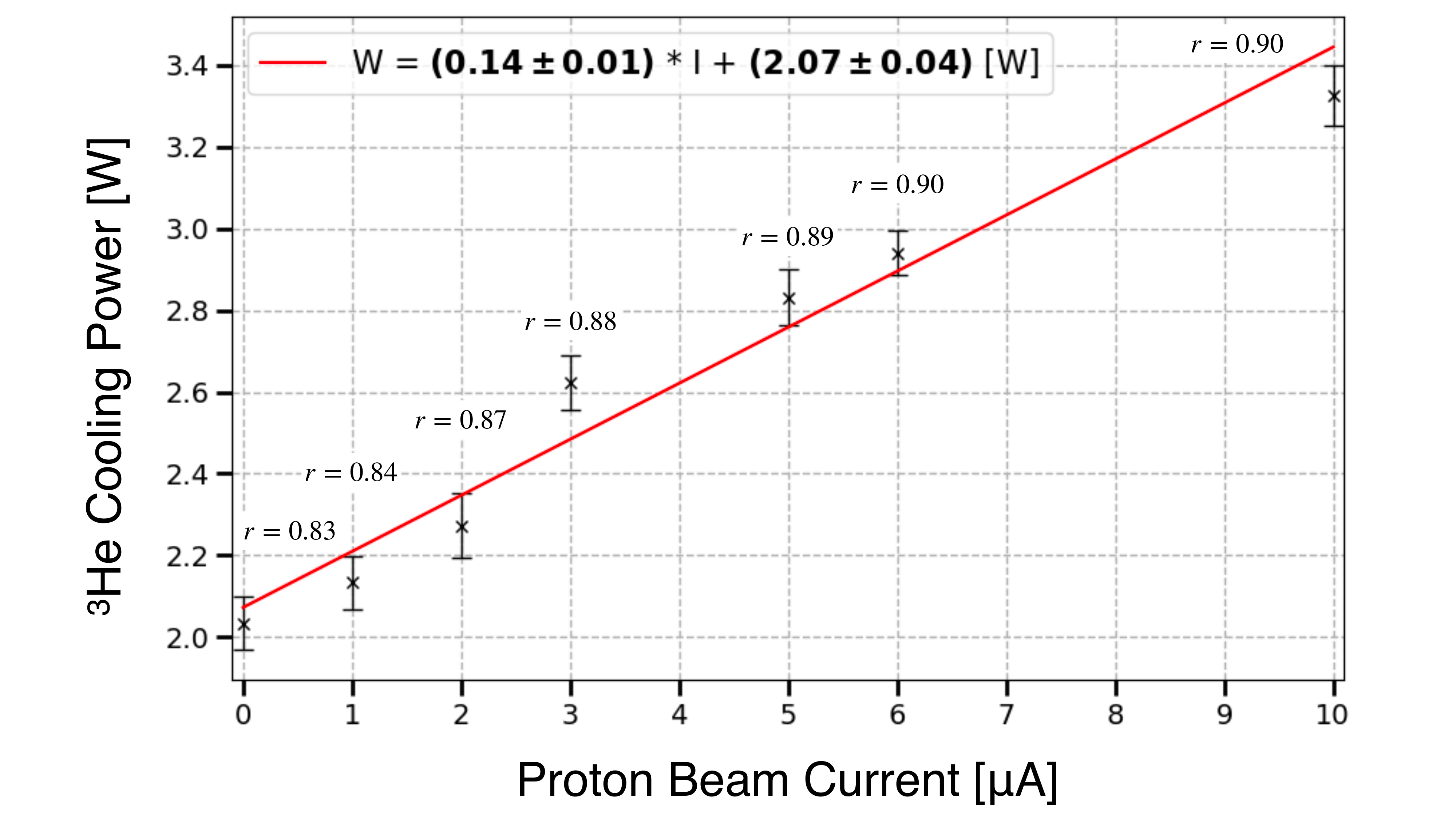}
\caption{Measurement of heat removed by $^3$He pumping as a function
  of beam current delivered to the spallation target, at a $^3$He
  temperature of 0.9~K; $r$ is the fraction of liquid $^3$He remaining
  after Joule-Thomson expansion.}
\label{fig:beam-heating}
\end{center}
\end{figure}

In experiments up to November 2024, the source was cooled and filled
with superfluid $^4$He.  During this time, the LD$_2$ cryostat was not
in place.  The cryogenic performance of the source was excellent.  A
base temperature of 0.8~K was achieved with an acceptable resting heat
load.  No evidence of a superleak was seen.  No clogs of either the
$^3$He or natural-abundance helium systems were experienced in $>$20
days of operation.  The temperatures in the tail section were
consistent with the superfluid helium being maintained at ${\sim}1$~K.
The beam heat load was measured (see Fig.~\ref{fig:beam-heating}) and
was consistent with expectations based on MCNP simulations within
10\%.  There was some evidence of a rise of temperature sensors closer
to the UCN production volume under the highest beam heat load,
consistent with our expectations based on heat conduction in turbulent
He-II (the Gorter-Mellink regime).

\begin{figure}[h]
\begin{center}
\includegraphics[width=\textwidth]{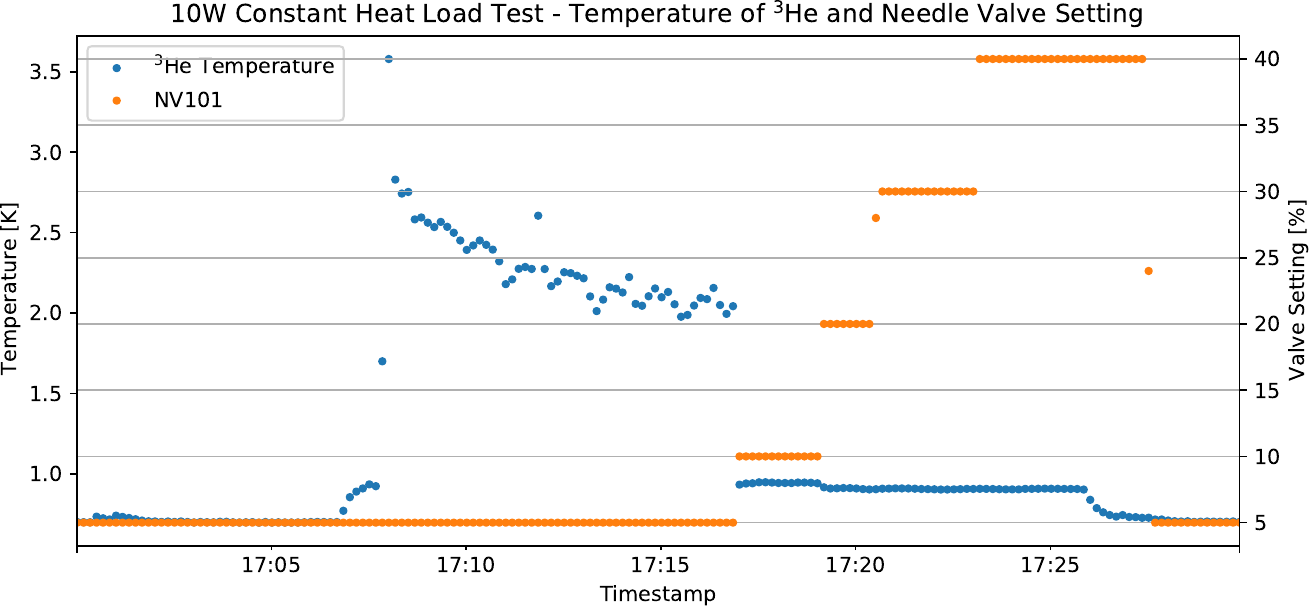}
\caption{Temperature of the $^3$He in the $^3$He-$^4$He copper heat
  exchanger, as a function of time during the application of 10~W of
  heat.  Blue points and left axis: $^3$He temperature.  Orange points
  and right axis: Joule-Thomson needle valve setting.  The $^3$He
  temperature rapidly reduces to 0.9~K when the needle valve is opened
  sufficiently.  Before and after the application of the 10~W of heat,
  the $^3$He temperature is 0.7~K.}
\label{fig:3he}
\end{center}
\end{figure}

The heat exchanger and refrigerator were capable of maintaining
sufficiently low $^4$He temperatures, in heater tests mimicking our
highest projected beam heat load of 10~W.  An example of this type of
test is presented in Fig.~\ref{fig:3he}, where the $^3$He pot
temperature is reduced to 0.9~K when the Joule-Thomson needle valve is
opened sufficiently.  During these tests, the $^3$He flow rate was
measured and, as in the beam heating tests, were found to be
consistent with expectation.

In runs where beam was present, we attempted to sense UCNs in a
detector connected directly to the UCN source (in place of the
superconducting polarizer in Fig.~\ref{fig:UCNsourceCAD}).  No
conclusive evidence of UCN detection was seen above the larger
background in this region, despite ${\sim}10^4$~UCNs/$\mu$A being
expected after saturation of the UCN density (a 60~s
irradiation time).  We strongly suspect this was caused by contamination of
the $^4$He production volume with either air or water frozen on the
inner surface.  As we were condensing $^4$He, we experienced clogging
of the condensation route in the cryostat, which necessitated filling
inefficiently through the recovery line.  This means that contaminants
could freeze on the coldest parts of our cryostat, in the
production volume.



Prior to our next runs, planned for June 2025, we plan to purify the
$^4$He using both our UCN source cryostat, and a new purifier system
from Japan, thus preventing future clogs.  This will allow us to fill
the cryostat in the proper way that it was designed to operate.  First
detection of UCN produced by the new source will be the key milestone
that we aim to achieve in the next runs.  This is a very exciting time
to be operating the facility, as we turn on the UCN source for the
first time and begin to verify its capabilities.

\section{Future Plans and Conclusion}
\label{conclusion}

The TUCAN project has made incredible progress in the past year, with
the cryogenic commissioning of the superfluid helium UCN source now
complete.  The next major installation for the project is the liquid
deuterium (LD$_2$) cryostat.  This is scheduled for spring 2025, and
is needed to boost UCN production by a factor of 30, which will make
the UCN source truly world-class.  The collaboration aims to complete
the milestones of first UCN detection, commissioning of the LD$_2$
cryostat, and delivery of UCN into the magnetically shielded room, by
the end of the calendar year in 2025.

In 2026, the TRIUMF laboratory will undergo a year-long shutdown in
support of the ARIEL project.  In 2027, the schedule calls for
commissioning of the TUCAN EDM experiment and preparation for data
taking.

Simultaneously with these efforts, we are applying for funding for an
upgrade to our helium liquefier facility, from the Canada Foundation
for Innovation and funding sources in Japan.  Included in the upgrade
are further improvements to the UCN source for UCN production and
delivery, improvements to the nEDM experiment to implement the
two-measurement-cell system shown in Fig.~\ref{fig:MesonCAD}, and
funding for further research into dual-species Xe-Hg comagnetometry.
These developments would be completed in 2028 and beyond.

\section{Acknowledgments}

We gratefully acknowledge the support of the Canada Foundation for
Innovation; the Canada Research Chairs program; the Natural Sciences
and Engineering Research Council of Canada (NSERC) SAPPJ-2016-00024,
SAPPJ-2019-00031, and SAPPJ-2023-00029; JSPS KAKENHI (Grant
Nos. 18H05230, 19K23442, 20KK0069, 20K14487, and 22H01236); JSPS
Bilateral Program (Grant No. JSPSBP120239940); JST FOREST Program
(Grant No.  JPMJFR2237); International Joint Research Promotion
Program of Osaka University; RCNP COREnet; the Yamada Science
Foundation; the Murata Science Foundation; the Grant for Overseas
Research by the Division of Graduate Studies (DoGS) of Kyoto
University; and the Universidad Nacional Aut\'onoma de M\'exico -
DGAPA program PASPA and grant PAPIIT AG102023.

\end{document}